# Quantum Communication with Continuum Single-Photon Pulses

F. F. S. Rios, R. V. Ramos*

*Lab. of Quantum Information Technology, Department of Teleinformatic Engineering – Federal University of Ceara - Fortaleza-Ce, Brazil.*
*Corresponding author: rubens.viana@pq.cnpq.br*

In this work, we analyze the behavior of continuum single-photon pulses in some quantum communication schemes. In particular, we consider the single-photon interference in a Mach-Zenhder interferometer, the HOM interference and the quantum bit commitment protocol.

Several protocols for quantum communication have been proposed however, in general their descriptions are based on single-frequency optical pulses. For example, single-photon interference is a crucial tool for experimental realization of quantum communication, quantum computation and quantum metrology schemes. Usually, the single-photon interference is analyzed as if the photons were produced by single-frequency optical sources, their propagation in optical fibers was free of dispersive effects and the behavior of the optical devices, mainly beam splitters and phase modulators, were not frequency-dependent. These are, obviously, simplifications of the real situation. A single-frequency source would violate the Heisenberg uncertainty principle and, hence, it does not exist. For example, the photon propagation in dispersive optical fibers, results in an additional phase term of the type

$$\exp\left[i\left(\beta L+(1/2)\beta_2\omega^2 L-(1/6)\beta_3\omega^3 L\right)\right] \quad (1)$$

where $\beta$ is the constant of propagation, $\beta_2$ is the GVD dispersion, $\beta_3$ is the third order dispersion (usually considered when $\beta_2 \sim 0$) and $L$ is the fiber's length propagated. Furthermore, real beam splitters, polarization rotators and phase modulators are frequency-dependent devices. Hence, a more realistic analysis of quantum communication schemes requires the consideration of continuum fields [1]. In this picture, the single-photon continuum state is given by

$$|1_\omega\rangle = \int_0^\infty \sigma(\omega)\hat{a}^+(\omega)d\omega|0_\omega\rangle \quad (2)$$

$$\int_0^\infty |\sigma(\omega)|^2 d\omega = 1. \quad (3)$$

The state $|0_\omega\rangle$ is the continuum vacuum state and, hence, $\hat{a}(\omega)|0_\omega\rangle = 0$, where $\hat{a}(\omega)$ is the continuum annihilation operator. At last, $|\sigma(\omega)|^2 d\omega$ gives the probability of the frequency of the photon to belong to the interval ($\omega$, $\omega + d\omega$).

In order to work with the continuum single-photon in quantum communication schemes, we firstly make its discretization. Let us start by writing $\sigma(\omega)$ in the basis of sinc functions:

$$\sigma(\omega) = \sum_{k=-\infty}^{\infty} \sigma(k\omega_s)\text{sinc}\left[(\omega-k\omega_s)/\omega_s\right] \quad (4)$$

$$\text{sinc}(x) = \sin(\pi x)/\pi x. \quad (5)$$

Using the orthogonality of the sinc function,

$$\frac{1}{\omega_s}\int_{-\infty}^{\infty}\text{sinc}\left[\frac{(\omega-k\omega_s)}{\omega_s}\right]\text{sinc}\left[\frac{(\omega-m\omega_s)}{\omega_s}\right]d\omega = \delta_{km}, \quad (6)$$

and the fact that $\sigma(\omega)$ is zero for negative frequencies, one has that

$$\int_0^\infty |\sigma(\omega)|^2 d\omega = \int_{-\infty}^{\infty} |\sigma(\omega)|^2 d\omega = \sum_{k=1}^{\infty}|\sigma(k\omega_s)|^2 \omega_s = 1. \quad (7)$$

Equation (7) shows us how to make discrete the continuum single-photon state:

$$|1_\omega\rangle = \sum_{k=1}^{\infty}\sigma(k\omega_s)\sqrt{\omega_s}|0\rangle_1 \otimes ... \otimes |1\rangle_k \otimes ... \quad (8)$$

According to (8) the continuum single-photon state can be approximated by a superposition of the tensor product of discrete oscillators. Each discrete oscillator is a mode that works in a single-frequency. For example, the state $|0\rangle_1\otimes...\otimes|1\rangle_k\otimes...$ means one photon in the frequency $k\omega_s$ and zero photons in the other frequencies. The number of

discrete oscillators is equal to the number of samples of $\sigma(\omega)$ and the amplitude of probability of $k$-th term in the superposition is given by $\sigma(k\omega_s)(\omega_s)^{1/2}$. Now, if $\sigma(\omega)$ vanishes for $\omega > N\omega_s$, then one has just a finite number of modes

$$|1_\omega\rangle = \sum_{k=1}^{N} \sigma(k\omega_s)\sqrt{\omega_s}|\tilde{1}\rangle_k \qquad (9)$$

$$|\tilde{1}\rangle_k = |0\rangle_1 \otimes ... \otimes |1\rangle_k \otimes ... \otimes |0\rangle_N. \qquad (10)$$

Now, we will consider the behavior of the quantum state given in (9) in a Mach-Zehnder interferometer (MZI) whose phase modulators are frequency-dependent (to include the frequency dependence of the beam splitters is just an algebra exercise). The MZI is composed by two lossless beam-splitters having transmittance $T = 1/2^{1/2}$ (and reflectance $R = i1/2^{1/2}$), and one phase modulator in each arm, $\phi_A(\omega)$ and $\phi_B(\omega)$. Such interferometer is useful in quantum key distribution (QKD) setups. The input state is $|1_\omega\rangle|0_\omega\rangle$. After some algebra one gets the following total quantum state at the interferometer output

$$|\psi_\omega\rangle = \sum_{k=1}^{N} \sigma(k\omega_s)\sqrt{\omega_s}|0\rangle_1|0\rangle_1 \otimes ... \otimes |\xi_\omega\rangle_k \otimes ... \otimes |0\rangle_N|0\rangle_N \qquad (11)$$

$$|\xi_\omega\rangle_k = ie^{i\Omega_k}\left\{\cos(\Delta_k)|1\rangle_k^a|0\rangle_k^b + \sin(\Delta_k)|0\rangle_k^a|1\rangle_k^b\right\} \qquad (12)$$

$$\Omega_k = [\phi_A(k\omega_s)+\phi_B(k\omega_s)]/2; \Delta_k = [\phi_A(k\omega_s)-\phi_B(k\omega_s)]/2. \qquad (13)$$

Hence, the probabilities of the photon to emerge at each output of the interferometer are

$$p_a = \sum_{k=1}^{N} \cos^2\left[\frac{\phi_A(k\omega_s)-\phi_B(k\omega_s)}{2}\right]|\sigma(k\omega_s)|^2 \omega_s \qquad (14)$$

$$p_b = \sum_{k=1}^{N} \sin^2\left[\frac{\phi_A(k\omega_s)-\phi_B(k\omega_s)}{2}\right]|\sigma(k\omega_s)|^2 \omega_s \qquad (15)$$

or, returning to the continuous case,

$$p_a = \int_0^\infty \cos^2\left[\frac{\phi_A(\omega)-\phi_B(\omega)}{2}\right]|\sigma(\omega)|^2 d\omega \qquad (16)$$

$$p_b = \int_0^\infty \sin^2\left[\frac{\phi_A(\omega)-\phi_B(\omega)}{2}\right]|\sigma(\omega)|^2 d\omega. \qquad (17)$$

Observing (16)-(17) one see that the frequency-dependence can increase the error rate of the QKD protocol (this error can be taken into account through the visibility of the interferometer) or it can be designedly used to increase the security of the protocol.

Now, let us consider the interference between two continuum single-photon pulses, coming from different single-photon sources, impinging in a beam splitter at the same time. The total state at the beam splitter's output is

$$U_{BS}|1_\omega\rangle|1_\omega\rangle = U_{BS}\left(\sum_{k=1}^{N}\sigma(k\omega_s)\sqrt{\omega_s}|\tilde{1}\rangle_k\right)\otimes\left(\sum_{l=1}^{N}\xi(l\omega_s)\sqrt{\omega_s}|\tilde{1}\rangle_l\right)$$

$$= \sum_{k,l=1}^{N}\sigma(k\omega_s)\xi(l\omega_s)\omega_s U_{BS}|\tilde{1}\rangle_k|\tilde{1}\rangle_l. \qquad (18)$$

In (18) $U_{BS}(\omega)$ is the unitary operation of the beam splitter. Its transmittance and reflectance are, respectively, $\cos(\theta(\omega))$ and $i\sin(\theta(\omega))$. Thus,

$$U_{BS}|\tilde{1}\rangle_k|\tilde{1}\rangle_l = \begin{cases} I) \text{ if } k \neq l \\ II) \text{ if } k = l \end{cases} \qquad (19)$$

$$I) \sum_{k=1}^{N}|0\rangle_1^a|0\rangle_1^b \otimes ... \otimes |\mu\rangle_k \otimes ... \otimes |\mu\rangle_l \otimes ... \otimes |0\rangle_N^a|0\rangle_N^b \qquad (20)$$

$$II) \sum_{k=1}^{N}|0\rangle_1^a|0\rangle_1^b \otimes ... \otimes |\lambda\rangle_k \otimes ... \otimes |0\rangle_N^a|0\rangle_N^b \qquad (21)$$

$$|\mu\rangle_k = \cos(\theta(k\omega_s))|1\rangle_k^a|0\rangle_k^b + i\sin(\theta(k\omega_s))|0\rangle_k^a|1\rangle_k^b \qquad (22)$$

$$|\lambda\rangle_k = \sin(2\theta(k\omega_s))\frac{|2\rangle_k^a|0\rangle_k^b + |0\rangle_k^a|2\rangle_k^b}{\sqrt{2}} + i\cos(2\theta(k\omega_s)) \qquad (23)$$

Using (18) – (23), one gets the coincidence probability

$$p_{coin} = \sum_{\substack{k,l=1 \\ k\neq l}}^{N} |\sigma(k\omega_s)\xi(l\omega_s)|^2 \omega_s^2 \begin{bmatrix}\cos^2(\theta(k\omega_s))\cos^2(\theta(l\omega_s))+\\ \sin^2(\theta(k\omega_s))\sin^2(\theta(l\omega_s))\end{bmatrix}$$

$$+\sum_{k=1}^{N}|\sigma(k\omega_s)\xi(k\omega_s)|^2 \omega_s^2 \cos^2(2\theta(k\omega_s)). \qquad (24)$$

If the beam splitter is not frequency-dependent and balanced ($\theta = \pi/4$) (24) reduces to

$$p_{coin} = \frac{1}{2} - \frac{1}{2}\sum_{k=1}^{N}|\sigma(k\omega_s)\xi(k\omega_s)|^2 \omega_s^2. \qquad (25)$$

At last, let us consider the quantum bit commitment protocol (QBC). It has been shown that QBC protocols cannot be unconditionally secure [2]. This is still a controversial question and some attempts of producing unconditional QBC protocols have been proposed [3,4]. Here, we consider the Lo-Chau's QBC protocol (LC-QBC) from a practical point of view, aiming to show that, at least in principle, Alice's cheating strategy may be noticed by Bob. The practical conditions considered are: the entangled photons have a spectral distribution and the quantum gates are frequency-dependent. The LC-QBC protocol can be explained in the following way: Alice and Bob agree that the states $|0_L\rangle = (|00\rangle+|11\rangle)/2^{1/2}$ and $|1_L\rangle = (|01\rangle+|10\rangle)/2^{1/2}$ represent, respectively, the logical bits '0' and '1'. In the commitment stage, Alice prepares the state $|0_L\rangle$ and she sends the second qubit to Bob. In the unveil stage two situations are possible: 1) Alice decides to keep the choice '0'. She measures her qubit in the $\{|0\rangle,|1\rangle\}$ basis and informs to Bob the values of the bit commited ('0') and the result of her measurement. Bob, by its turn, measures his qubit in the same basis and compares the result with that one announced by Alice. If the results of the measurements are the same, Bob thinks that Alice acted honestly. 2) Alice changes her mind and decides to unveil the value '1'. She applies the not gate $X$ and makes a measurement in her qubit. Alice informs to Bob the values of the bit commited ('1') and the result of her measurement. Bob, by its turn, measures his qubit and compares the result with that one announced by Alice. If the measurement results are different, Bob thinks that Alice acted honestly. Hence, since Alice can always change from '0' to '1' (by applying the $X$ gate in her qubit) without being noted, she can always cheating Bob with zero probability of being caught cheating. This scenario changes when we consider real entangled states. Let us consider that Alice and Bob will run the LC-QBC protocol using the following entangled state

$$|0_L\rangle = \int_0^\infty d\Omega \sigma(\Omega) \frac{|\omega_0+\Omega,\omega_0-\Omega\rangle_{HH} + |\omega_0+\Omega,\omega_0-\Omega\rangle_{VV}}{\sqrt{2}}. \qquad (26)$$

The discretization of the state (26) using (4)-(7) is

$$|0_L\rangle = \sum_{\substack{k,l=1 \\ k+l=M}}^{N} \sigma(k\omega_s, l\omega_s)\sqrt{\omega_s} \left[ \frac{|\tilde{1}\rangle_k^H |\tilde{1}\rangle_l^H + |\tilde{1}\rangle_k^V |\tilde{1}\rangle_l^V}{\sqrt{2}} \right]. \qquad (27)$$

$$M\omega_s = 2\omega_0. \qquad (28)$$

According to (27), with probability $|\sigma(k\omega_s,l\omega_s)|^2\omega_s$ the photons in the frequencies $k\omega_s$ and $l\omega_s$ ($k\omega_s + l\omega_s = M\omega_s = 2\omega_0$) are in the entangled state $(|HH\rangle+|VV\rangle)/2^{1/2}$. The not gate, by its turn, is a frequency-dependent polarization rotator. It is represented by $R[\theta(\omega)]$, where $\theta(\omega) = \pi/2$ in the central frequency. When Alice tries to cheat applying $R[\theta(\omega)]$, she produces the quantum state

$$R[\theta(\omega)]|0_L\rangle = \sum_{\substack{k,l=1 \\ k+l=M}}^{N} \sigma(k\omega_s, l\omega_s)\sqrt{\omega_s}$$

$$\times \left[ \frac{\cos[\theta(k\omega_s)]|\tilde{1}\rangle_k^H|\tilde{1}\rangle_l^H + \sin[\theta(k\omega_s)]|\tilde{1}\rangle_k^V|\tilde{1}\rangle_l^H}{\sqrt{2}} + \sin[\theta(k\omega_s)]|\tilde{1}\rangle_k^H|\tilde{1}\rangle_l^V - \cos[\theta(k\omega_s)]|\tilde{1}\rangle_k^V|\tilde{1}\rangle_l^V \right]. \qquad (29)$$

An error in Bob denouncing Alice's cheating strategy will occur when Alice informs that she chose bit '1' and Bob gets in his measurement the same result as Alice got in her measurement. Using the state in (29) one gets the following error probability

$$PE = \sum_{\substack{k,l=1 \\ k+l=M}}^{N} |\sigma(k\omega_s, l\omega_s)|^2 \cos[\theta(k\omega_s)]\omega_s \qquad (30)$$

or, returning to the continuum case,

$$PE = \int_0^\infty d\Omega |\sigma(\Omega)|^2 \cos^2[\theta(\omega_0+\Omega)]. \qquad (31)$$

In (31) it is assumed that Alice (Bob) kept the photon with frequency $\omega_0+\Omega$ ($\omega_0-\Omega$). Hence, once one takes into account the spectral distribution and the frequency dependence, we may note that Alice's strategy may cause an error in Bob.

In conclusion, this work showed that considering the spectral distribution of single-photons is an important issue in the analysis of quantum error rate and security of quantum communication protocols. On the other hand, the discretization of the continuum single-photon using the sinc functions makes easy the calculation of the important probabilities considered in the error rate and security analysis.

This work was supported by the Brazilian agency CNPq Grant no. 303514/2008-6. Also, this work was performed as part of the Brazilian National Institute of Science and Technology for Quantum Information.

.